\documentclass[prc,showpacs,eqsecnum,preprint,nofootinbib]{revtex4}
\usepackage{amsfonts,amsmath,amssymb,bm,graphicx,float}
\newcommand{\ds}{\displaystyle}
\begin{document}

\title[]{Bound-state beta-decay of a neutron in a strong magnetic field}%
\author{Konstantin A. Kouzakov}%
\affiliation{Department of Nuclear Physics and Quantum
Theory of Collisions, Moscow State University, Moscow 119992, Russia}%
\email{kouzakov@srd.sinp.msu.ru}%
\author{Alexander I. Studenikin}%
\affiliation{Department of Theoretical Physics,
Moscow State University, Moscow 119992, Russia}%
\email{studenik@srd.sinp.msu.ru}

\begin{abstract}The beta-decay of a neutron into a bound $(pe^-)$ state
and an antineutrino in the presence of a strong uniform magnetic
field ($B\gtrsim10^{13}$~G) is considered. The beta-decay process
is treated within the framework of the standard model of weak
interactions. A Bethe-Salpeter formalism is employed for
description of the bound $(pe^-)$ system in a strong magnetic
field. For the field strengths $10^{13}$~G$\lesssim
B\lesssim10^{18}$~G the estimate for the ratio of the bound-state
decay rate $w_b$ and the usual (continuum-state) decay rate $w_c$
is derived. It is found that in such strong magnetic fields
$w_b/w_c \sim 0.1 \div 0.4$. This is in contrast to the field-free
case, where $w_b/w_c \simeq 4.2 \times 10^{-6}$ [J.~N.~Bahcall,
Phys. Rev. {\bf 124}, 495 (1961); L.~L.~Nemenov, Sov. J. Nucl.
Phys. {\bf 15}, 582 (1972); X.~Song, J. Phys. G: Nucl. Phys. {\bf
13}, 1023 (1987)]. The dependence of the ratio $w_b/w_c$ on the
magnetic field strength $B$ exhibits a logarithmic-like behavior.
The obtained results can be important for applications in
astrophysics and cosmology.
\end{abstract}

\date{\today}

\maketitle

\section{Introduction}
\label{intro}
It is well known that besides the main decay mode of a free
neutron in vacuum
\begin{equation}
\label{beta_decay_free} n\rightarrow p+e^-+\bar{\nu}_e
\end{equation}
there is also the neutron decay into a bound $(pe^-)$ state
(hydrogen atom) and an antineutrino
\begin{equation}
\label{beta_decay_bound} n\rightarrow(pe^-)+\bar{\nu}_e.
\end{equation}
Several theoretical estimates for the rate of the latter process
have been performed (see Refs.~\cite{bahcall61,nemenov71,song87}
and references therein). Regardless of the framework employed for
the description of the hydrogen atom (for example, such as the
Schr\"odinger equation~\cite{nemenov71}, Dirac
theory~\cite{bahcall61} or relativistic Bethe-Salpeter
formalism~\cite{song87}) all the treatments yield the following
estimate for the ratio of bound-state and continuum-state decay
rates: $w_b/w_c\simeq 4.2\times10^{-6}$. Therefore, in general
one might expect the effect of the neutron bound-state
decay~(\ref{beta_decay_bound}) to be subdominant if compared with
the continuum-state decay~(\ref{beta_decay_free}). However, as
shown below this conclusion does not hold true in the situation
where a neutron decays in the presence of a strong magnetic field.

In this work we estimate the ratio $w_b/w_c$ in the presence of a
magnetic field with strength $B\gtrsim10^{13}$~G. Our study is
motivated by a widely accepted view that in diverse astrophysical
and cosmological environments the physics of neutrinos in strong
magnetic fields plays an important role. The existence of very
strong magnetic fields in proto-neutron stars and pulsars is well
established. The surface magnetic fields of many supernovae and
neutron stars are of the order of
$B\sim10^{12}\div10^{14}$~G~\cite{fushiki92,duncan92}. The
surface magnetic fields of magnetars are perhaps as large as
$B\sim10^{15}\div10^{16}$~G~\cite{kulkarni98}. Very strong
magnetic fields are also supposed to have existed in the early
Universe (see Ref.~\cite{grasso01} for a recent review).

The effect of a constant magnetic field on the
process~(\ref{beta_decay_free}) is well documented. In short, the
constant magnetic field affects the motion of a charged particle
in such a way that the energy associated with the transverse
motion (with respect to the field direction) is quantized into
Landau levels while the longitudinal motion remains free. The
larger the field intensity $B$, the larger the energy separating
the Landau levels. Thus, the decay rate $w_c$ exhibits the
following dependence on the field intensity $B$. For $0<B\ll
B_{cr}$, where $B_{cr}=(\Delta^2-m_e^2)/2e=1.2\times10^{14}$~G
with $\Delta=m_n-m_p$ being the nucleon mass defect (hereafter we
use the units $\hbar=c=1$), the effect of a magnetic field on
both an electron and a proton is small and $w_c$ remains
practically insensitive to $B$. As the field intensity approaches
the value $B_{cr}$ (note that $B_{cr}>B_e$, where
$B_e=m_e^2/e=4.414\times10^{13}$~G is the so-called Schwinger
field) the effect of Landau quantization on the transverse
electron motion becomes considerable $\omega_c\sim
B$~\cite{korovina64}. If $B_{cr}<B\ll B_{cr}'$, where
$B_{cr}'=[(m_n-m_e)^2-m_p^2]/2e=1.25\times10^{17}$~G, the
electron (due to the energy conservation law) can occupy only the
lowest Landau level, and it was shown in~\cite{korovina64} that
the decay rate grows as $w_c\sim B$. In the case of a superstrong
magnetic field ($B\geq B_{cr}'$) the proton can occupy only the
lowest Landau level and one has again a monotonic dependence
$w_c\sim B$~\cite{zakhartsev85}. Finally, as the field strength
exceeds the value $B\sim 1.5\times10^{18}$~G the modification of
the strong forces, which bind the quarks in nucleons, is such
that it can close the mass gap between the neutron and proton
making the neutron stable~\cite{bander92} (see
also~\cite{grasso01}). In what follows we concentrate on field
strengths in the range $B_e\lesssim B\ll B_p$, where
$B_p=m_p^2/e=1.5\times10^{20}$~G is the Schwinger field for the
proton.

To our knowledge, no theoretical analysis has been published for
the process~(\ref{beta_decay_bound}) in a (strong) magnetic
field. It should be noted that such study is hampered by the fact
that in calculating the decay rate $w_b$ one must know the
eigenenergies and eigenstates of the hydrogen atom in an external
magnetic field. In the case of a strong field the latter problem
has been studied in a number of papers, where numerical and
semianalytical results have been derived within different
approaches based on the
Schr\"odinger~\cite{shiff39,loudon59,eliott60,hasegawa61,rau76,kashiev80,wang95,potekhin01,rutkowski03,karnakov03},
Dirac~\cite{lindgren79,goldman91,chen93}, and Breit~\cite{doman80}
equations. A common result of all the treatments is that, compared
to the field-free case, the ground state of the hydrogen atom is
(1) very tightly bound and (2) well separated in energy from
excited states. The shape of the atom is also affected by the
field. The transverse atomic size is approximately determined by
the magnetic length $a_B=1/\sqrt{eB}\ll a_0$, where
$a_0=(e^2m_e)^{-1}$ is the Bohr radius, and the longitudinal
atomic size $a_\parallel$ depends on the binding atomic energy. In
particular, in the case $B\gtrsim B_{cr}$ one has for the ground
state a cigar-like shape with $a_\parallel\ll a_0$, while for the
excited states $a_\parallel\sim a_0$~\cite{karnakov03}. It should
be noted that these remarkable effects of a strong magnetic field
on the hydrogen atom are obtained assuming the absence of
transverse motion of the atom as a whole. However, even at a
moderate transverse atomic velocity a rather strong electric field
is induced in the rest frame of the atom pushing the electron and
the proton apart. As a result, the binding energy decreases as the
transverse velocity increases. Thus, in contrast to the field-free
case, the center-of-mass atomic motion does not decouple from the
relative internal motion, and therefore a correct description of
the hydrogen atom in a strong magnetic field should involve the
two-body approach~\cite{herold81,potekhin94_98}.

By analogy with the field-free case~\cite{bahcall61}, the ratio
$w_b/w_c$ can be roughly estimated using a phase-space argument
that does not depend on a formal theory of weak interactions. The
phase-space volume available for the final products in the
process~(\ref{beta_decay_free}) can be presented by the function
$f(B,\Delta)$ that depends on the field strength and nucleon mass
defect. For the bound-state decay~(\ref{beta_decay_bound}), the
available phase-space volume is given by
$\varepsilon_\nu^2|\psi(0)|^2$, where $\varepsilon_\nu$ is the
neutrino's energy and $\psi(0)$ is the ground-state wave function
of a hydrogen atom evaluated at zero distance between an electron
and a proton. The ratio of the bound-state and continuum-state
decay rates is approximately equal to the ratio of the
corresponding phase-space volumes
\begin{equation}
\label{estimate_phspace}
\frac{w_b}{w_c}\sim\frac{\varepsilon_\nu^2|\psi(0)|^2}{f(B,\Delta)}.
\end{equation}
For example, using the phase-space argument in the field-free case
($B=0$) we get the result
\begin{equation}
\label{estimate_phspace_free} \frac{w_b}{w_c}\sim\pi
e^6\left(\frac{\Delta}{m_e}-1\right)^2\approx2.9\times10^{-6},
\end{equation}
which is of the same order of magnitude as the accurate
calculations of Refs.~\cite{bahcall61,nemenov71,song87}. In the
case of a strong field ($B_{cr}<B\ll B_{cr}'$) we have
$$
f(B,\Delta)\sim eBm_e^3/\pi^2~\textrm{and}~|\psi(0)|^2\sim (\pi
a_\parallel a_B^2)^{-1}=(\pi a_\parallel/eB)^{-1}.
$$
Using the estimates for $a_\parallel$ reported
in~Ref.\cite{karnakov03}, we might expect [in accordance with
Eq.~(\ref{estimate_phspace})] that
\begin{equation}
\label{estimate_phspace_bound} \frac{w_b}{w_c}\sim\frac{\pi
e^2a_0}{a_\parallel}\left(\frac{\Delta}{m_e}-1\right)^2\gtrsim0.1.
\end{equation}
This result is almost five orders of magnitude larger than in the
field-free case~(\ref{estimate_phspace_free}) and thus it calls
for a more detailed and rigorous theoretical analysis, such as
carried out below.

The paper is organized as follows. In Sec.~\ref{gen_form} a
general formulation of the problem is given in the context of the
standard quantum field theory employing a Bethe-Salpeter
formalism. Then, in Sec.~\ref{wf}, we discuss in detail a
structure of the wave function of a bound $(pe^-)$ state in a
strong magnetic field. Further, specific approximations to the
Bethe-Salpeter equation are developed in the cases $B_{cr}<B\ll
B_{cr}'$ and $B_{cr}'<B\ll B_p$. Sec.~\ref{estimate} is devoted to
the derivation of an estimate for the bound-state decay rate. In
addition, the asymptotic formula for the ratio of bound-state and
continuum-state decay rates is obtained. The conclusions are drawn
in Sec.~\ref{conclude}.

\section{General theory}
\label{gen_form}
In the case of field strengths relevant to this work
($B_e\lesssim B\ll B_p$) we can neglect the effect of a magnetic
field on both the propagator of $W$ boson and the weak form
factors. Therefore, the transition matrix element for the decay
process~(\ref{beta_decay_bound}) can be written as follows
\begin{equation}
\label{matr_el} T_{fi}=\frac{G}{\sqrt{2}}\int\bar{\chi}^E(x,x)
\gamma_{\mu}(1+\alpha\gamma_5)\psi_n
\gamma^{\mu}(1+\gamma_5)\psi_{\nu}d^4x,
\end{equation}
where $\psi_n$ and $\psi_{\nu}$ are the Dirac wave functions of
the neutron and neutrino, respectively, $G=G_F\cos\theta_c$,
$\theta_c$ is the Cabibbo angle, and $\alpha=1.26$ is the ratio
of the axial and vector constants. The conjugate $\bar{\chi}^E$
of the wave function $\chi^E$ describing the bound $(pe^-)$
system with energy $E$ is determined as
$\bar{\chi}^E=\gamma^0\chi^{E\dag}\gamma^0$. In the ladder
approximation the wave function $\chi^E$ satisfies the following
Bethe-Salpeter equation
%
%
\begin{equation}
\label{bethe-salpeter} \chi^E(x_e,x_p)= ie^2\int
G^e(x_e,x_e')\gamma^\mu G^p(x_p,x_p')\gamma^\nu
D_{\mu\nu}(x_e',x_p')\chi^E(x_e',x_p')d^4x_e'd^4x_p'.
\end{equation}
Here $D_{\mu\nu}(x,x')$ is the photon propagator. The electron
and proton propagators satisfy the Dirac equation in an external
magnetic field
\begin{equation}
\label{dirac} [\gamma^\mu(-i\partial_\mu\mp
eA_\mu)+m_{e/p}]G^{e/p}(x,x')=\delta^{(4)}(x-x').
\end{equation}

In what follows we use the symmetric axial gauge of the vector
potential
\begin{equation}
\label{ax_guage} {\bf A}({\bf r})=\frac{1}{2}[{\bf B}\times{\bf
r}].
\end{equation}
The photon propagator is supposed to be invariant under the time
and space translations
\begin{equation}
\label{photon_inv} D_{\mu\nu}(x,x')=D_{\mu\nu}({\bf r}-{\bf
r}',t-t'),
\end{equation}
whereas the electron and proton propagators are invariant under
the time translation
\begin{equation}
\label{time_trans_inv} G^{e/p}(x,x')=G^{e/p}({\bf r},{\bf
r}',t-t').
\end{equation}
Their properties with respect to the space translation are
determined according to Eqs.~(\ref{dirac}) and~(\ref{ax_guage}) as
follows
\begin{equation}
\label{trans_inv} G^{e/p}({\bf r}-{\bf r}_C,{\bf r}'-{\bf
r}_C,t-t')=\exp\left\{\pm i\frac{e}{2}[{\bf B}\times{\bf
r}_C]({\bf r}-{\bf r}')\right\}G^{e/p}({\bf r},{\bf r}',t-t').
\end{equation}
From~(\ref{photon_inv}),~(\ref{trans_inv}),
and~(\ref{bethe-salpeter}) we receive
\begin{equation}
\label{time_tr_inv_bs}\chi^E(x_e,x_p)=\exp(-iET)\exp\left\{i\left({\bf
P}+\frac{e}{2}[{\bf B}\times{\bf r}]\right){\bf R}\right\}
\eta^{E,{\bf P}}({\bf r},t),
\end{equation}
where ${\bf P}$ is the total pseudomomentum which corresponds to
the total momentum in the field-free case~\cite{herold81},
$t=t_e-t_p$ and ${\bf r}={\bf r}_e-{\bf r}_p$ are respectively the
relative internal time and space position, while $T$ and ${\bf R}$
refer to the center-of-mass time and space coordinates.

Thus, passing in~(\ref{bethe-salpeter}) to Fourier-transform
according to the rule
\begin{equation}
\label{fourier-trans} G^{e/p}({\bf r},{\bf r}',t-t')=\int
G^{e/p}({\bf r},{\bf
r}',\varepsilon)\exp[-i\varepsilon(t-t')]\frac{d\varepsilon}{2\pi},
\end{equation}
we obtain the equation
\begin{eqnarray}
\label{bethe-salpeter_1} \eta^{E,{\bf P}}({\bf r},t)=
\frac{ie^2}{2\pi}\int\exp[-i\varepsilon(t-t')] G^e({\bf r},{\bf
R}'+{\bf r}',\varepsilon)\gamma^\mu
G^p(0,{\bf R}',E-\varepsilon)\gamma^\nu D_{\mu\nu}({\bf r}',t') \nonumber\\
\times \exp\left\{i\left({\bf P}+\frac{e}{2}[{\bf B}\times{\bf
r}']\right){\bf R}'\right\}\eta^{E,{\bf P}}({\bf
r}',t')d\varepsilon d{\bf r}'d{\bf R}'dt'.\nonumber\\
\end{eqnarray}

We will assume that the vector potential~(\ref{ax_guage}) is
specified in the rest frame of the neutron, which is supposed to
be not affected by a magnetic field. Thus, the neutron wave
function is given by
%
%
\begin{equation}
\label{neutron_wf}\psi_n=u_{n}e^{-im_nt}, \quad
u_{n}=\left(\begin{array}{c}w_n\\0\end{array}\right) \quad
(w_n^*w_n=1),
\end{equation}
where $p_n=(m_n,0)$, $w_n$ is the two-component neutron spinor.
The neutrino is supposed to be massless and not interacting with a
magnetic field. Therefore its wave function can be presented as
follows
\begin{equation}
\label{neutrino_wf}\psi_\nu=u_{\nu} e^{-i(\varepsilon_\nu t-{\bf
p}_\nu{\bf r})}, \quad
u_{\nu}=\frac{1}{\sqrt{2}}\left(\begin{array}{c}v_\nu\\-v_\nu\end{array}\right)
\quad (v_\nu^*v_\nu=1),
\end{equation}
where $p_\nu=(\varepsilon_\nu,{\bf p}_\nu)$,
$\varepsilon_\nu=|{\bf p}_\nu|$, and $v_\nu$ is the two-component
eigenspinor of the spin-projection operator, $({\bm\sigma}{\bf
p}_\nu)v_\nu=-\varepsilon_\nu v_\nu$.

Accounting for Eqs.~(\ref{time_tr_inv_bs}),~(\ref{neutron_wf}),
and~(\ref{neutrino_wf}), the matrix element~(\ref{matr_el}) takes
the form
\begin{eqnarray}
\label{matr_el_1}
T_{fi}=\frac{G}{\sqrt{2}}(2\pi)^4\delta(m_n-\varepsilon_\nu-E)\delta^{(3)}({\bf
p}_\nu-{\bf P})
\bar{\eta}^{E,{\bf P}}(0,0)\gamma_{\mu}(1+\alpha\gamma_5)u_{n}
\gamma^{\mu}(1+\gamma_5)u_{\nu}.
\end{eqnarray}
Using the rules~\footnote{The normalization length and time are
set to unity.} $|2\pi\delta(E)|^2=2\pi\delta(E)$ and
$|(2\pi)^3\delta^{(3)}({\bf p})|^2=(2\pi)^3\delta^{(3)}({\bf p})$,
we get for the rate $w_b$ of the decay
process~(\ref{beta_decay_bound})
\begin{equation}
\label{decay_rate} w_b=\frac{G^2}{2}\sum\nolimits^{(pe^-)}\int
(2\pi)^4\delta(m_n-\varepsilon_\nu-E)\delta^{(3)}({\bf p}_\nu-{\bf
P}) |\mathcal{M}|^2\frac{d{\bf p}_\nu}{(2\pi)^3}\frac{d{\bf
P}}{(2\pi)^3},
\end{equation}
where
\begin{equation}
\label{matr_el_2} \mathcal{M}=\bar{\eta}^{E,{\bf
P}}(0,0)\gamma_{\mu}(1+\alpha\gamma_5)u_{n}
\gamma^{\mu}(1+\gamma_5)u_{\nu}
\end{equation}
is a reduced matrix element, the sum $\sum\nolimits^{(pe^-)}$ runs
over all bound states of the ($pe^-$) system and the average over
neutron spin states is assumed.

%
\section{The Bethe-Salpeter wave function}
\label{wf}
The key element that determines the decay rate~(\ref{decay_rate})
is the wave function of the bound ($pe^-$) system. We solve the
Bethe-Salpeter equation~(\ref{bethe-salpeter_1}) neglecting the
retardation effects~\cite{karplus52} (this amounts to the
non-relativity of internal motion), so that the photon propagator
in the Coulomb gauge assumes the form
\begin{equation}
\label{photon_propagator} D_{\mu\nu}({\bf
r},t)=-\delta_\mu^0\delta_\nu^0\frac{\delta(t)}{r}.
\end{equation}
Inserting~(\ref{photon_propagator}) in the Bethe-Salpeter
equation~(\ref{bethe-salpeter_1}), we obtain
\begin{eqnarray}
\label{bethe-salpeter_2} \eta^{E,{\bf P}}({\bf r},0)=
-\frac{ie^2}{2\pi}\int G^e({\bf r},{\bf R}'+{\bf
r}',\varepsilon)\gamma^0 G^p(0,{\bf
R}',E-\varepsilon)\gamma^0\frac{1}{r'}\nonumber\\
\times\exp\left\{i\left({\bf P}+\frac{e}{2}[{\bf B}\times{\bf
r}']\right){\bf R}'\right\}\eta^{E,{\bf P}}({\bf
r}',0)d\varepsilon d{\bf r}'d{\vec R}'.
\end{eqnarray}
To carry out an integration over $\varepsilon$ in the kernel of
this equation, we expand the electron and proton propagators into
series over Dirac eigenstates in a magnetic field keeping only the
positive energy pole contributions~\cite{shabad86}:
\begin{equation}
\label{green_function} G^{e/p}({\bf r},{\bf r}',\varepsilon) =
\sum_\kappa\frac{\psi^{(\kappa)}_{e/p}({\bf
r})\bar{\psi}^{(\kappa)}_{{e/p}}({\bf
r}')}{\varepsilon-\varepsilon_{\kappa}+i0}.
\end{equation}
If $z$ axis is directed along the magnetic field vector ${\bf B}$,
then $\varepsilon_{\kappa}=\sqrt{m_{e/p}^2+p_z^2+2neB}$ is the
energy of the Dirac eigenstate specified by a set of the quantum
numbers $\kappa=\{n,p_z,j_z,s\}$, where the discrete numbers
$n=0,1,2,\ldots$ denote the Landau levels, $p_z$ is the
longitudinal momentum, $j_z$ is the $z$ component of the total
angular momentum, and $s=\pm1$ is equivalent to the spin quantum
number in the field-free case (see, for
instance,~\cite{sokolov68}). In cylindrical space coordinates, the
Dirac wave functions in Eq.~(\ref{green_function}) can be
presented as the products of longitudinal and transverse parts
\begin{equation}
\label{dirac_spinor}\psi^{(\kappa)}_{e/p}({\bf
r})=\exp(ip_zz)\lambda^{(\tilde{\kappa})}_{e/p}(p_z,{\bm
\rho})\quad ({\bm \rho}\equiv{\bf r}_\perp),
\end{equation}
where $\lambda^{(\tilde{\kappa})}_{e/p}(p_z,{\bm \rho})$ is the
transverse spinor labeled by a set of the quantum numbers
$\tilde{\kappa}=\{n,j_z,s\}$. The explicit forms of the
transverse electron and proton spinors are
\begin{eqnarray}
\label{transv_spinor}\lambda^{(n,j_z,+1)}_{e}(p_z,{\bm
\rho})=\frac{1}{\sqrt{2\varepsilon_{\kappa_e}}}
\left(\begin{array}{c}\sqrt{m_e+\varepsilon_{\kappa_e}}\phi_{n-1,j_z-1/2}({\bm \rho})\\0\\
\ds\frac{p_z}{\sqrt{m_e+\varepsilon_{\kappa_e}}}\phi_{n-1,j_z-1/2}({\bm \rho})\\
\ds
i\sqrt{\frac{2neB}{m_e+\varepsilon_{\kappa_e}}}\phi_{n,j_z+1/2}({\bm
\rho})\end{array}\right),
\nonumber\\
\lambda^{(n,j_z,-1)}_{e}(p_z,{\bm
\rho})=\frac{1}{\sqrt{2\varepsilon_{\kappa_e}}}
\left(\begin{array}{c}0\\\sqrt{m_e+\varepsilon_{\kappa_e}}\phi_{n,j_z+1/2}({\bm \rho})\\
\ds-i\sqrt{\frac{2neB}{m_e+\varepsilon_{\kappa_e}}}\phi_{n-1,j_z-1/2}({\bm \rho})\\
\ds\frac{-p_z}{\sqrt{m_e+\varepsilon_{\kappa_e}}}\phi_{n,j_z+1/2}({\bm
\rho})\end{array}\right),
\nonumber\\
\lambda^{(n,j_z,+1)}_{p}(p_z,{\bm
\rho})=\frac{1}{\sqrt{2\varepsilon_{\kappa_p}}}
\left(\begin{array}{c}\sqrt{m_p+\varepsilon_{\kappa_p}}\phi_{n,j_z-1/2}({\bm \rho})\\0\\
\ds\frac{p_z}{\sqrt{m_p+\varepsilon_{\kappa_p}}}\phi_{n,j_z-1/2}({\bm
\rho})\\
\ds-i\sqrt{\frac{2neB}{m_p+\varepsilon_{\kappa_p}}}\phi_{n-1,j_z+1/2}({\bm
\rho})\end{array}\right),
\nonumber\\
\lambda^{(n,j_z,-1)}_{p}(p_z,{\bm
\rho})=\frac{1}{\sqrt{2\varepsilon_{\kappa_p}}}
\left(\begin{array}{c}0\\\sqrt{m_p+\varepsilon_{\kappa_p}}\phi_{n-1,j_z+1/2}({\bm \rho})\\
\ds
i\sqrt{\frac{2neB}{m_p+\varepsilon_{\kappa_p}}}\phi_{n,j_z-1/2}({\bm
\rho})\\
\ds\frac{-p_z}{\sqrt{m_p+\varepsilon_{\kappa_p}}}\phi_{n-1,j_z+1/2}({\bm
\rho})\end{array}\right),
\end{eqnarray}
where the transverse Landau orbitals are given
by~\cite{lindgren79,sokolov68}
\begin{eqnarray}
\label{landau_orbital} \phi_{n_\rho,m}({\bm
\rho})=\sqrt{\frac{n_\rho!}{2\pi
a_B^2(n_\rho-m)!}}\exp(im\varphi)\exp\left(-\frac{\rho^2}{4a_B^2}\right)
\left(\frac{\rho^2}{2a_B^2}\right)^{-m/2}
L_{n_\rho}^{(-m)}\left(\frac{\rho^2}{2a_B^2}\right)\nonumber\\
(n_\rho=0,1,2,\ldots,\quad n_\rho-m=0,1,2,\ldots)
\end{eqnarray}
with $L_{n_\rho}^{(-m)}$ being a generalized Lagguerre polynomial.

Using Eqs.~(\ref{green_function}) and~(\ref{dirac_spinor}), we
have
\begin{equation}
\label{green_function_1} G^{e/p}({\bf r},{\bf r}',\varepsilon)
=\frac{1}{2\pi}\sum_{\tilde\kappa}\int
\frac{\lambda^{(\tilde\kappa)}_{e/p}(p_z,{\bm
\rho})\bar{\lambda}^{(\tilde\kappa)}_{e/p}(p_z,{\bm
\rho}')}{\varepsilon-\sqrt{\varepsilon_{\tilde\kappa}^2+p_z^2}+i0}
\exp[ip_z(z-z')]dp_z,
\end{equation}
where $\varepsilon_{\tilde\kappa}=\sqrt{m_{e/p}^2+2neB}$ is the
energy of the transverse motion.
Inserting~(\ref{green_function_1}) in~(\ref{bethe-salpeter_2})
and integrating over $\varepsilon$, we receive
\begin{eqnarray}
\label{bethe-salpeter_4} \eta^{E,{\bf P}}({\bf r},0)=
-\frac{e^2}{2\pi}\sum_{\tilde\kappa_e,\tilde\kappa_p}
\int\frac{\exp[iq_z(z-z')]\lambda^{(\tilde\kappa_e)}_{e}(q_z,{\bm
\rho})\lambda^{(\tilde\kappa_p)}_{p}(P_z-q_z,0)}
{E-\sqrt{\varepsilon_{\tilde\kappa_e}^2+q_z^2}
-\sqrt{\varepsilon_{\tilde\kappa_p}^2+(P_z-q_z)^2}}
\lambda^{(\tilde\kappa_e)*}_{e}(q_z,{\bm\rho}'+{\bf R}_\perp')
\nonumber\\ \times\lambda^{(\tilde\kappa_p)*}_{p}(P_z-q_z,{\bf
R}_\perp')
 \frac{1}{\sqrt{\rho'^2+z'^2}} \exp\left\{i\left({\bf
P}_\perp+\frac{e}{2}[{\bf B}\times{\bm \rho}']\right){\bf
R}_\perp'\right\} \nonumber\\ \times\eta^{E,{\bf P}}({\bf r}',0)
dq_z d{\bm \rho}'dz'd{\bf R}_\perp',
\end{eqnarray}
where the sum over $\tilde\kappa_p$ involves only the transverse
proton spinors with $j_z=\pm1/2$ [according to
Eqs.~(\ref{transv_spinor}) and~(\ref{landau_orbital}) the
transverse proton spinors with $j_z\neq\pm1/2$ vanish at ${\bm
\rho}=0$].

We consider below the following two cases of field strengths: (1)
$B_{cr}<B\ll B_{cr}'$ and (2) $B_{cr}'<B\ll B_p$. For each of
these cases we develop the corresponding approximation to the
Bethe-Salpeter equation~(\ref{bethe-salpeter_4}) taking into
account the kinematical regime of the decay
process~(\ref{beta_decay_bound}).

\subsection{The case $B_{cr}<B\ll B_{cr}'$}
\label{wf_1}
Recall that in the case $B>B_{cr}$ the energy conservation law
dictates that the electron in the usual decay
process~(\ref{beta_decay_free}) can occupy only the lowest Landau
level. Considering the bound-state decay
process~(\ref{beta_decay_bound}) in the case $B>B_{cr}$, we note
that [as follows from Eq.~(\ref{bethe-salpeter_4})] the electron
can (virtually) occupy not only the lowest Landau level. However,
the probability for the electron to occupy the excited Landau
levels is vanishingly small due to the following fact. Virtual
electron transitions between the lowest and excited Landau levels
are induced by the Coulomb electron-proton interaction, which in
the present case of field strengths is much weaker than the
interaction of the electron with a magnetic field. Thus, the
electron contribution from excited Landau levels to the wave
function of the bound $(pe^-)$ system is negligible. Therefore,
one can leave in Eq.~(\ref{bethe-salpeter_4}) only those electron
terms that correspond to the lowest Landau level (this amounts to
the adiabatic approximation). In accordance with
Eq.~(\ref{transv_spinor}), the transverse electron spinors for
the lowest Landau level ($n=0$) are given by
\begin{eqnarray}
\label{zero_landau_transv_spinor}\lambda^{(0,j_z,-1)}_{e}(p_z,{\bm
\rho})=u(p_z)\phi_{0,m}({\bm \rho}), \quad
m=j_z+\frac{1}{2}=0,-1,-2,\ldots,\nonumber\\ u(p_z)=
\left(\begin{array}{c}0\\\ds\sqrt{\frac{m_e+\varepsilon_{\kappa_e}}{2\varepsilon_{\kappa_e}}}\\
0\\
\ds\frac{-p_z}{\sqrt{{2\varepsilon_{\kappa_e}}(m_e+\varepsilon_{\kappa_e})}}\end{array}\right),
\end{eqnarray}
where $\varepsilon_{\kappa_e}=\sqrt{m_e^2+p_z^2}$. In addition,
the non-relativistic approximation applies to the proton
transverse spinors and energies:
\begin{eqnarray}
\label{proton_spinor_approx} \lambda^{(n,1/2,+1)}_{p}(p_z,{\bm
\rho}) =u_p^{(+)}\phi_{n,0}(\rho), \quad
\lambda^{(n,-1/2,-1)}_{p}(p_z,{\bm \rho})
=u_p^{(-)}\phi_{n-1,0}(\rho)~[\phi_{-1,0}(\rho)\equiv0],
\nonumber\\
\varepsilon_{\kappa_p}=m_p+n\omega_p+\frac{p_z^2}{2m_p},\quad
u_p^{(+)}=\left(\begin{array}{c}1\\0\\0\\0\end{array}\right),
\quad
u_p^{(-)}=\left(\begin{array}{c}0\\1\\0\\0\end{array}\right),
\end{eqnarray}
where $\omega_p=eB/m_p$ is the Larmor frequency for the proton.

Taking into account Eqs.~(\ref{zero_landau_transv_spinor})
and~(\ref{proton_spinor_approx}) and using
$\phi_{n,0}(0)=\sqrt{eB/2\pi}$, we arrive at
\begin{equation}
\label{intermediate} \eta^{E_\pm,{\bf P}}({\bf r},0)=f({\bf
r})u_p^{(\pm)},
\end{equation}
where $E_-=E_+ +\omega_p$ and
\begin{eqnarray}
\label{bethe-salpeter_5} f_{k}({\bf r})=
-\frac{e^2}{2\pi}\sqrt{\frac{eB}{2\pi}} \sum_{n,m}\phi_{0,m}({\bm
\rho}) \int\frac{\exp[iq_z(z-z')]u_k(q_z)u_l(q_z)}
{E_+-E_e(q_z)-E_p(n,P_z-q_z)} \phi^*_{0,m}({\bm\rho}'+{\bf
R}_\perp') \phi_{n,0}(R_\perp')\nonumber\\
\times\frac{1}{\sqrt{\rho'^2+z'^2}}
  \exp\left\{i\left({\bf
P}_\perp+\frac{e}{2}[{\bf B}\times{\bm \rho}']\right){\bf
R}_\perp'\right\}f_{l}({\bf r}') dq_z d{\bm
\rho}'dz'd{\bf R}_\perp'.\nonumber\\
\end{eqnarray}
Here
$$
E_e(q_z)=\sqrt{m_e^2+q_z^2} \text{~and~}
E_p(n,P_z-q_z)=m_p+n\omega_p+\frac{(P_z-q_z)^2}{2m_p},
$$
$k,l=1,2,3,4$ stand for the spinor indices and a sum over the
repeating spinor index $l$ is assumed. To carry out an integration
over ${\bf R}_\perp'$ in the kernel of
Eq.~(\ref{bethe-salpeter_5}), we use an expansion of the
undisplaced Landau state over an infinite set of the displaced
Landau orbitals~\cite{sokolov68}
\begin{eqnarray}
\label{landau_orbital_expand} \phi_{n,m}({\bm
\rho})=\exp\left(i\frac{e}{2}[{\bf B}\times{\bm \rho}]{\bm
\rho}_0\right)
\sqrt{\frac{2\pi}{eB}}\sum_{m'}\phi_{n-m',m-m'}({\bm\rho}_0)
\phi_{n,m'}({\bm \rho}-{\bm\rho}_0).
\end{eqnarray}
Involving the following integral
\begin{eqnarray}
\label{factor} \int \exp\left(i{\bf P}_\perp{\bm
\rho}\right)\phi_{n-m',m-m'}^*({\bm\rho})\phi_{n-N,0}(\rho) d{\bm
\rho}=\frac{2\pi}{eB}\phi_{n-m',N-m'}^*({\bm\rho}_C)
\phi_{n-m,N-m}({\bm\rho}_C),
\end{eqnarray}
where ${\bm \rho}_C=e[{\bf P}_\perp\times{\bf B}]/(eB)^2$, and
introducing
\begin{equation}
\label{displaced_state} g({\bf r})=\exp\left(-\frac{i}{2}{\bf
P}_\perp{\bf r}\right) f({\bf r}-{\bm \rho}_C),
\end{equation}
we get
\begin{eqnarray}
\label{bethe-salpeter_6} g_{k}({\bf r})=
-\frac{e^2}{2\pi}\sum_{m}\phi_{0,m}({\bm \rho})
\int\frac{\exp[iq_z(z-z')]u_k(q_z)u_l(q_z)}
{E_+-E_e(q_z)-E_p(|m|,P_z-q_z)}\phi^*_{0,m}({\bm\rho}')
\nonumber\\
\times\frac{1}{\sqrt{({\bm \rho}'-{\bm
\rho}_C)^2+z'^2}}g_{l}({\bf r}') dq_z d{\bm
\rho}'dz'.
\end{eqnarray}
Let us seek a solution to this equation in a form of the following
expansion
\begin{equation}
\label{solution_expand} g({\bf r})=\sum_{m} h(m;z)\phi_{0,m}({\bm
\rho}).
\end{equation}
Inserting it in Eq.~(\ref{bethe-salpeter_6}) we obtain an infinite
set of the coupled integral equations
\begin{eqnarray}
\label{bethe-salpeter_7} h_{k}(m;z)= \frac{1}{2\pi} \sum_{m'}
\int\frac{\exp[iq_z(z-z')]u_k(q_z)u_l(q_z)}
{E_+-E_e(q_z)-E_p(|m|,P_z-q_z)}
V_{mm'}({\bm \rho}_C,z')h_{l}(m';z')dq_z dz'.
\end{eqnarray}
Here the one-dimensional potential functions $V_{mm'}({\bm
\rho}_C,z)$ are given by
\begin{equation}
\label{effective_1d_potential} V_{mm'}({\bm
\rho}_C,z)=-e^2\int\frac{\phi_{0,m}^*({\bm \rho})\phi_{0,m'}({\bm
\rho})}{\sqrt{({\bm \rho}-{\bm \rho}_C)^2+z^2}}d{\bm \rho}.
\end{equation}
Examining their properties, we deduce that in the case ${\bf
P}_\perp=0$ the infinite set of integral
equations~(\ref{bethe-salpeter_7}) becomes decoupled
\begin{eqnarray}
\label{bethe-salpeter_8} h_{k}(m;z)= \frac{1}{2\pi}
\int\frac{\exp[iq_z(z-z')]u_k(q_z)u_l(q_z)}
{E_+-E_e(q_z)-E_p(|m|,P_z-q_z)} 
V_{m}(z')h_{l}(m;z')dq_z dz',
\end{eqnarray}
where
\begin{eqnarray}
\label{effective_1d_potential_1}
V_m(z)=-e^2\int\frac{|\phi_{0,m}({\bm
\rho})|^2}{\sqrt{\rho^2+z^2}}d{\bm \rho}=
-e^2\exp\left(\frac{z^2}{4a_B^2}\right)
\left(\frac{z^2}{2a_B^2}\right)^{\frac{|m|}{2}-\frac{1}{4}}
W_{-\frac{|m|}{2}-\frac{1}{4},\frac{|m|}{2}+\frac{1}{4}}\left(\frac{z^2}{2a_B^2}\right)
\nonumber\\
\end{eqnarray}
with $W_{\mu,\nu}$ being a Whittaker function~\cite{abramowitz72}.

In the non-relativistic limit, the
equation~(\ref{bethe-salpeter_8}) is equivalent to the
one-dimensional Schr\"odinger equation for a hydrogen atom in the
adiabatic approximation
\begin{equation}
\label{schrodinger}
\left[\frac{1}{2\mu}\frac{d^2}{dz^2}-V_m(z)+E^\parallel_{m}\right]\psi_{m}(z)=0,
\end{equation}
where $\mu=(m_e+m_p)/2m_em_p$ is the reduced mass and
\begin{equation}
\label{schrodinger_en_wf}
E^\parallel_{m}=E_+-m_p-m_e-\frac{P_z^2}{2(m_e+m_p)}-|m|\omega_p,
\quad \psi_{m}(z)=\exp\left(\frac{iP_zm_ez}{m_p}\right)h_{2}(m;z).
\end{equation}
Here it should be pointed out that the use of the non-relativistic
approximation~(\ref{schrodinger}) for the treatment of the bound
states in the case of a magnetic field with strength $B\gtrsim
B_{cr}$ can be justified for the following reasons~\cite{lai01}:
(1) the shape of the transverse electron wave function determined
by the Landau orbital~(\ref{landau_orbital}) in the relativistic
theory is the same as in the non-relativistic theory; (2) the
electron remains non-relativistic in the $z$ direction as long as
the binding energy $|E^\parallel|\ll m_e$.

The one-dimensional Schr\"odinger equation~(\ref{schrodinger}) has
been studied in~\cite{loudon59,eliott60,hasegawa61,karnakov03}.
Here we outline its basic properties that are important for our
purposes. The eigensolutions of Eq.~(\ref{schrodinger}) can be
categorized in two distinct classes: (1) the states having no node
and (2) the states having node(s) in their wave functions. The
states having no node in their wave functions are tightly bound.
For each $m$ there is one such state, which is the most tightly
bound (the ground state) in the case $m=0$. The ground-state
energy $E^\parallel\ll-13.6$~eV, i.e. it is much lower than that
of a hydrogen atom in the field-free case. On the contrary, the
states having nodes in their wave functions are weakly bound. For
example, in the case $m=0$ the state with one node in its wave
function has about the same energy as the ground state of a
hydrogen atom in the field-free case, $E^\parallel\simeq-13.6$~eV,
whereas the states with more than one node in their wave functions
have the higher energies ($E^\parallel>-13.6$~eV).

\subsection{The case $B_{cr}'<B\ll B_p$}
\label{wf_2}

When considering the influence of very strong magnetic fields on
the neutron beta-decay, one should be careful about the effect of
a magnetic field on the anomalous magnetic moments of the neutron
and proton. In particular, it is known (see, for instance,
Ref.~\cite{grasso01}) that the interplay between the anomalous
magnetic moments of the neutron and proton shifts the energies of
these particles making the neutron stable in the case
$B\sim1.5\times10^{18}$~G. However, if the field strength is
$B\lesssim 1.5\times10^{18}$~G the effect of the nucleon anomalous
magnetic moments is still subdominant. Note that the
corresponding shift of the electron energy due to the electron
anomalous magnetic moment vanishes in such a strong
field~\cite{TerBagDorJETP69, Schw88} (see also Ref.~\cite
{grasso01}). In what follows, we neglect the effects connected
with the anomalous magnetic moments, assuming that the field
strength does not exceed the value $B\sim1.5\times10^{18}$~G.

In the case of a superstrong magnetic field ($B_{cr}'<B\ll B_p$)
both the electron and the proton in the usual decay
process~(\ref{beta_decay_free}) can occupy only the lowest Landau
level. Using the same arguments as in the preceding subsection
with regard to the electron's dynamics in the bound-state decay
process~(\ref{beta_decay_bound}), we again arrive at
Eq.~(\ref{bethe-salpeter_7}). However, noticing that the
(virtual) proton transitions between Landau levels induced by the
Coulomb electron-proton interaction are almost completely
suppressed, we can neglect the couplings between different values
of $m$ in Eq.~(\ref{bethe-salpeter_7}) and consider only the case
$m=0$. Thereby, instead of the infinite set of coupled
equations~(\ref{bethe-salpeter_7}), we obtain the single integral
equation
\begin{eqnarray}
\label{bethe-salpeter_9} h_{k}(0;z)=\frac{1}{2\pi}
\int\frac{\exp[iq_z(z-z')]u_k(q_z)u_l(q_z)}
{E_+-E_e(q_z)-E_p(0,P_z-q_z)}
V_{00}({\bm \rho}_C,z')h_{l}(0;z')dq_z dz'.
\end{eqnarray}
The corresponding Schr\"odinger equation is
\begin{equation}
\label{schrodinger_1}
\left[\frac{1}{2\mu}\frac{d^2}{dz^2}-V_{00}({\bm
\rho}_C,z)+E^\parallel_{0}({\bm \rho}_C)\right]\psi_{0}({\bm
\rho}_C,z)=0,
\end{equation}
where ${\bm \rho}_C$ plays a role of parameter and
\begin{equation}
\label{schrodinger_en_wf_1} E^\parallel_{0}({\bm
\rho}_C)=E_+-m_p-m_e-\frac{P_z^2}{2(m_e+m_p)}, \quad \psi_{0}({\bm
\rho}_C,z)=\exp\left(\frac{iP_zm_ez}{m_p}\right)h_{2}(0;z).
\end{equation}
Note that if $\rho_C\ll a_B$, we have
\begin{equation}
\label{approx_potential} V_{00}({\bm \rho}_C,z)\approx
V_{0}(z)=-\frac{e^2}{a_B}\sqrt{\frac{\pi}{2}}{\rm
erfc}\left(\frac{|z|}{\sqrt{2}a_B}\right)
\exp\left(\frac{z^2}{2a_B^2}\right)=\left\{\begin{array}{ll}
\displaystyle
-\sqrt{\frac{\pi}{2}}\frac{e^2}{a_B},& \quad |z|\rightarrow0,\\
\ds -\frac{e^2}{|z|},& \quad
|z|\rightarrow\infty,\end{array}\right.
\end{equation}
where ${\rm erfc}(y)$ is the complementary error
function~\cite{abramowitz72}.

\section{The decay rate}
\label{estimate}
In this section, using the results of Sec.~\ref{wf} for the wave
function of the bound $(pe^-)$ system, we derive the bound-state
decay rates in the particular cases, namely, $B_{cr}<B\ll B_{cr}'$
and $B_{cr}'<B\ll B_p$. Then we utilize the obtained bound-state
decay rates in the extrapolation procedure, when treating the
general case $B_e\lesssim B\ll B_p$.

\subsection{The case $B_{cr}<B\ll B_{cr}'$}
\label{estimate_1}
Using Eqs.~(\ref{intermediate}),~(\ref{displaced_state}),
and~(\ref{solution_expand}) we perform an integration over ${\bf
P}$ in Eq.~(\ref{decay_rate}) and obtain for the bound-state decay
rate
\begin{eqnarray}
\label{decay_rate_1} w_b=\frac{G^2}{2}\sum_\tau\int
2\pi\left\{\delta[m_n-\varepsilon_\nu-E_{+,\tau}({\bf p}_\nu)]
|\mathcal{M}_{+,\tau}({\bf
p}_\nu)|^2\right.\nonumber\\
\left. +\delta[m_n-\varepsilon_\nu-E_{-,\tau}({\bf
p}_\nu)]|\mathcal{M}_{-,\tau}({\bf p}_\nu)|^2\right\}\frac{d{\bf
p}_\nu}{(2\pi)^3}.
\end{eqnarray}
Here the indices $\tau$ label the solutions of
Eq.~(\ref{bethe-salpeter_7}) with corresponding eigenenergies
$E_{\pm,\tau}({\bf p}_\nu)$ [$E_{-,\tau}({\bf
p}_\nu)=E_{+,\tau}({\bf p}_\nu)+\omega_p$]. In accordance with
Eq.~(\ref{matr_el_2}), the reduced matrix elements are given by
\begin{equation}
\label{matr_el_3} \mathcal{M}_{{\pm},\tau}({\bf
p}_\nu)=\sum_{m}\phi_{0,m}^*({\bm
\rho}_\nu)[\bar{u}_p^{(\pm)}\gamma_{\mu}(1+\alpha\gamma_5)u_{n}]
[\bar{h}_{\tau}(m;0)\gamma^{\mu}(1+\gamma_5)u_{\nu}],
\end{equation}
where ${\bm \rho}_\nu=e[{\bf p}_{\nu}\times{\bf B}]/(eB)^2$.

To evaluate the reduced matrix element~(\ref{matr_el_3}), we note
that for the field strengths under consideration one has
$\rho_\nu^{max}/a_B=p_\nu^{max} a_B\lesssim1$, where $p_\nu^{max}$
is the maximal possible value for the neutrino's momentum in the
process~(\ref{beta_decay_bound}). This allows us to put
$\rho_\nu=0$ in Eq.~(\ref{matr_el_3}), so that we get
\begin{eqnarray}
\label{decay_rate_2} w_b=\frac{eBG^2}{2}\sum_\tau\int
\left\{\delta[\Delta-\varepsilon_\nu-m_e-E^\parallel_\tau(p_{\nu,z})]
|\mathcal{M}_{+,\tau}({\bf p}_\nu)|^2 \right.\nonumber\\
\left.
+\delta[\Delta-\varepsilon_\nu-m_e-\omega_p-E^\parallel_\tau(p_{\nu,z})]
|\mathcal{M}_{-,\tau}({\bf p}_\nu)|^2\right\} \frac{d{\bf
p}_\nu}{(2\pi)^3}.
\end{eqnarray}
Here
\begin{eqnarray}
\label{matr_el_4} \mathcal{M}_{+,\tau}({\bf p}_\nu)&=&
\sqrt{2}[h^{p_{\nu,z}*}_{\tau,2}(0)-h^{p_{\nu,z}*}_{\tau,4}(0)]
[(1+\alpha)\omega_{n,1}v_{\nu,2}-2\alpha\omega_{n,2}v_{\nu,1}],\nonumber\\
\mathcal{M}_{-,\tau}({\bf p}_\nu)&=&
\sqrt{2}[h^{p_{\nu,z}*}_{\tau,2}(0)-h^{p_{\nu,z}*}_{\tau,4}(0)](1-\alpha)\omega_{n,2}v_{\nu,2},
\end{eqnarray}
where in accordance with Eq.~(\ref{bethe-salpeter_7}) the wave
functions $h^{p_{\nu,z}}_\tau$ are the solutions of the equation
\begin{eqnarray}
\label{bethe-salpeter_10} h^{p_{\nu,z}}_i(z)= \frac{1}{2\pi}
\int\frac{\exp[iq_z(z-z')]u_i(q_z)u_l(q_z)} {\ds
E^\parallel(p_{\nu,z})+m_e+m_p-E_e(q_z)-E_p(0,p_{\nu,z}-q_z)}
V_{0}(z')h^{p_{\nu,z}}_{l}(z')dq_z dz'.
\end{eqnarray}
In the non-relativistic approximation
\begin{equation}
\label{nonrelativistic_approx}
E^\parallel_\tau(p_{\nu,z})=E^\parallel_\tau+\frac{p_{\nu,z}^2}{2(m_e+m_p)},
\quad h^{p_{\nu,z}}_{\tau,2}(0)=\psi_\tau(0), \quad
h^{p_{\nu,z}}_{\tau,4}(0)=0,
\end{equation}
where $\psi_\tau$ are the solutions of the Schr\"odinger
equation~(\ref{schrodinger}) for $m=0$ with energies
$E^\parallel_{0,\tau}\equiv E^\parallel_\tau$. The index
$\tau=0,1,...$ corresponds to the number of nodes in the wave
function $\psi_\tau(z)$.

Using Eq.~(\ref{nonrelativistic_approx}) and neglecting $\omega_p$
and $p_\nu^2/2(m_e+m_p)$ with respect to $\Delta-m_e-E^\parallel$,
we obtain (after averaging over neutron spin states) the following
result for the bound-state decay rate:
\begin{equation}
\label{decay_rate_3}
w_b=\frac{eBG^2}{8\pi^2}(1+3\alpha^2)\sum_\tau|\psi_\tau(0)|^2(\Delta-m_e-E^\parallel_\tau)^2,
\end{equation}
where the sum runs over even states ($\tau=0,2,\ldots$). The odd
states ($\tau=1,3,\ldots$) do not contribute to the bound-state
decay rate because their wave functions vanish at the origin
[$\psi_\tau(0)\equiv0$].

Let us derive the ratio of the bound-state decay rate $w_b$ [see
Eq.~(\ref{decay_rate_3})] and the continuum-state decay rate
$w_c$, i.e. the decay rate of the usual
process~(\ref{beta_decay_free}). To facilitate an accurate
estimation of the ratio $w_b/w_c$, we calculate $w_c$ under the
same assumptions that have been made to obtain
Eq.~(\ref{decay_rate_3}), i.e. we assume $\rho_\nu=0$ and
$\omega_p,~p_\nu^2/2(m_e+m_p)\ll\Delta-m_e$. Thus, we get
\begin{equation}
\label{decay_rate_free}
\frac{w_b}{w_c}=\frac{\pi\sum_\tau|\psi_\tau(0)|^2}{Cm_e}
\left(\frac{\Delta-E^\parallel_\tau}{m_e^2}-1\right)^2,
\end{equation}
where
\begin{equation}
\label{factor_decay}
C=\frac{1}{3}\left(\frac{\Delta^2}{m_e^2}+2\right)\sqrt{\frac{\Delta^2}{m_e^2}-1}
-\frac{\Delta}{m_e}{\rm arccosh}\left(\frac{\Delta}{m_e}\right).
\end{equation}

It is useful to compare the result~(\ref{decay_rate_free}) with
that in the field-free case (see also
Refs.~\cite{bahcall61,nemenov71}):
\begin{equation}
\label{field_free}
\frac{w_b}{w_c}=\frac{2\pi^2\sum_n|\psi_n(0)|^2}{C_0m_e^3}
\left(\frac{\Delta-\varepsilon_n}{m_e}-1\right)^2,
\end{equation}
where
\begin{equation}
\label{factor_decay_field_free}
C_0=\frac{1}{60}\left(\frac{2\Delta^4}{m_e^4}-\frac{9\Delta^2}{m_e^2}-8\right)\sqrt{\frac{\Delta^2}{m_e^2}-1}
+\frac{\Delta}{4m_e}{\rm arccosh}\left(\frac{\Delta}{m_e}\right),
\end{equation}
$n=1,2,\ldots$ is the principal quantum number, $\psi_n$
[$|\psi_n(0)|^2=(4\pi n^3a_0^3)^{-1}$] and
$\varepsilon_n=-(2a_0n^2)^{-1}$ are respectively the wave
function and the binding energy of the $ns$ state of a hydrogen
atom. It is seen that both expressions~(\ref{decay_rate_free})
and~(\ref{field_free}) have a similar structure. However, the
important difference consists in that $\psi_\tau$ are the
one-dimensional wave functions while $\psi_n$ are the
three-dimensional ones. This difference is also reflected in the
appearance of the factor $m_e^{-1}$ in the right-hand side of
Eq.~(\ref{decay_rate_free}) instead of the factor $m_e^{-3}$
occurring in the field-free case~(\ref{field_free}).

\subsection{The case $B_{cr}'<B\ll B_p$}
\label{estimate_2}
By analogy with the previous subsection, in the case of a
magnetic field with strength in the range $B_p'<B\ll B_p$ we have
\begin{equation}
\label{decay_rate_10} w_b=\frac{G^2}{2}\sum_\tau\int
2\pi\delta[m_n-\varepsilon_\nu-E_{+,\tau}({\bf p}_\nu)]
|\mathcal{M}_{+,\tau}({\bf p}_\nu)|^2\frac{d{\bf
p}_\nu}{(2\pi)^3},
\end{equation}
where the reduced matrix elements are given by
\begin{equation}
\label{matr_el_30} \mathcal{M}_{+,\tau}({\bf
p}_\nu)=\sqrt{\frac{eB}{2\pi}}\exp\left(-\frac{p_{\nu,\perp}^2}{4eB}\right)
[\bar{u}_p^{(+)}\gamma_{\mu}(1+\alpha\gamma_5)u_{n}]
[\bar{h}_\tau(0;0)\gamma^{\mu}(1+\gamma_5)u_{\nu}]
\end{equation}
with $h_\tau$ being the solutions of Eq.~(\ref{bethe-salpeter_9}).
Following the approximate procedure developed in the preceding
subsection, we deduce that
\begin{equation}
\label{decay_rate_20}
w_b=\frac{eBG^2}{8\pi^2}(1+2\alpha+5\alpha^2)\sum_\tau|\psi_\tau(0)|^2(\Delta-m_e-E^\parallel_\tau)^2,
\end{equation}
where $\psi_\tau$ and $E^\parallel_\tau$ are specified
in~(\ref{nonrelativistic_approx}). And for the ratio of
bound-state and continuum-state decay rates we obtain
\begin{equation}
\label{decay_rate_free_1}
\frac{w_b}{w_c}=\frac{\pi\sum_\tau|\psi_\tau(0)|^2}{Cm_e}
\left(\frac{\Delta-E^\parallel_\tau}{m_e}-1\right)^2.
\end{equation}
It is remarkable that this expression is identical with the one
obtained in the case $B_{cr}<B\ll B_{cr}'$ [see
Eq.~(\ref{decay_rate_free})]. We utilize this fact below, when
estimating the ratio $w_b/w_c$ in the general case $B_e\lesssim
B\ll B_p$.

\subsection{The case $B_e\lesssim B\ll B_p$}
\label{estimate_3}
A smooth dependence of $w_c$ on the field strength
$B$~\cite{korovina64,zakhartsev85} combined with the identity of
the expressions~(\ref{decay_rate_free})
and~(\ref{decay_rate_free_1}) allows us to extrapolate the
obtained results for the ratio $w_b/w_c$ in the whole range of
considered field strengths. Specifically, we assume the
expression~(\ref{decay_rate_free}) [(\ref{decay_rate_free_1})]
obtained in the case $B_{cr}<B\ll B_{cr}'$ ($B_{cr}'<B\ll B_p$)
to be valid for estimating the ratio $w_b/w_c$ in the general
case $B_e\lesssim B\ll B_p$, which incorporates the above
particular ranges of field strengths.

Let us note that in the field-free case ($B=0$) the relative
contribution of excited states to the ratio $w_b/w_c$ [see
Eq.~(\ref{field_free})] is about 20\%~\cite{song87,bahcall61}.
Taking into account that, in contrast to the field-free situation,
in the presence of a strong magnetic field the excited states are
very weakly bound in comparison with the ground state, we neglect
their contribution to the ratio $w_b/w_c$. The ground-state wave
function $\psi(z)$ can be approximated as
follows~\cite{loudon59,leinson85}
\begin{equation}
\label{wf_approx}
\psi(z)=\left(2\mu|E^\parallel|\right)^{1/4}\exp\left(-\sqrt{2\mu|E^\parallel|}|z|\right),
\end{equation}
where $E^\parallel$ is the ground-state energy. Thus we get
($\mu\simeq m_e$)
\begin{equation}
\label{decay_rate_free_2}
\frac{w_b}{w_c}=\frac{\pi}{C}\sqrt{\frac{2|E^\parallel|}{m_e}}
\left(\frac{\Delta-E^\parallel}{m_e}-1\right)^2.
\end{equation}
In the field-free case the corresponding result is given by
(accounting only for the ground-state contribution)
\begin{equation}
\label{field_free_1}
\frac{w_b}{w_c}=\frac{\pi}{C_0}\sqrt{\frac{2|\varepsilon|}{m_e}}\frac{|\varepsilon|}{m_e}
\left(\frac{\Delta-\varepsilon}{m_e}-1\right)^2,
\end{equation}
where $\varepsilon=-e^4m_e/2$ is the ground-state energy of the
hydrogen atom.

\begin{figure}
\begin{center}
\includegraphics[width=16cm]{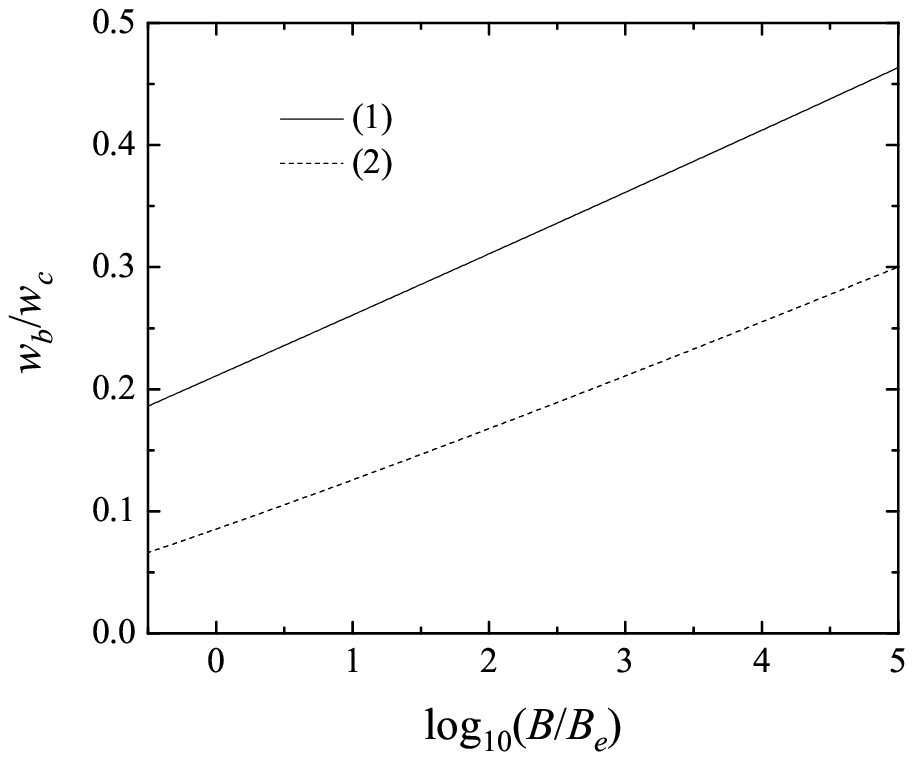}
\end{center}
\caption{\label{fig1}The estimate for the ratio of bound-state
and continuum-state decay rates in the case of a strong magnetic
field. The cases, where the asymptotic values for $E^\parallel$
are derived according to Ref.~\cite{landau77} [see Eq.~(\ref{LL})]
and Ref.~\cite{karnakov03} [see Eq.~(\ref{KP})], are represented
by the solid and dashed lines, respectively.}
\end{figure}

Fig.~\ref{fig1} displays the numerical results for the ratio
$w_b/w_c$ calculated in accordance with
Eq.~(\ref{decay_rate_free_2}) using two asymptotic formulas for
$E^\parallel$: (1) the well-known formula from
Ref.~\cite{landau77} (see also
Refs.~\cite{loudon59,eliott60,rau76})
%
\begin{equation}
\label{LL}
E^\parallel=\varepsilon\ln^2\left(\frac{eB}{2m_e|\varepsilon|}\right)
\end{equation}
and (2) the recent formula from Ref.~\cite{karnakov03}
\begin{equation}
\label{KP} E^\parallel=\varepsilon\ln^2\left[\frac{c_\infty
eB/2m_e|\varepsilon|}{\ln^2(eB/2m_e|\varepsilon|)}\right],
\end{equation}
where $c_\infty=0.2809$. It is seen that both asymptotic formulas
give the same order of magnitude for the ratio $w_b/w_c$. In both
cases the results exhibit a linear dependence on
$\log_{10}(B/B_e)$. However, the results obtained on a basis of
Eq.~(\ref{LL}) are approximately two times larger in magnitude
than those obtained on a basis of Eq.~(\ref{KP}). Due to the fact
that the asymptotic formula~(\ref{KP}) yields more accurate
values for the ground-state energy $E^\parallel$ than the
formula~(\ref{LL}) does (see Ref.~\cite{karnakov03} for details),
we can conclude that the estimate for the ratio $w_b/w_c$ using
Eq.~(\ref{KP}) is more realistic.

\section{Summary and conclusions}
\label{conclude}
In summary, we have considered and analyzed theoretically the
neutron decay into a bound $(pe^-)$ state and an antineutrino in
the presence of a magnetic field with strength
$B\gtrsim10^{13}$~G.
The amplitude of the bound-state decay process has been formulated
within the framework of the standard model of weak interactions.
For the description of the bound $(pe^-)$ system we have employed
the Bethe-Salpeter equation. The approximations to the
Bethe-Salpeter wave function of the bound $(pe^-)$ state have
been specified depending on the field strength $B$. We have
derived the bound-state decay rate $w_b$ in two particular cases,
namely, $B_{cr}<B\ll B_{cr}'$ and $B_{cr}'<B\ll B_p$. In both
cases, the identical expressions for the ratio of bound-state and
continuum-state decay rates $w_b/w_c$ have been obtained. We have
estimated the ratio $w_b/w_c$ in the general case $B_e\lesssim
B\ll B_p$ using two asymptotic
formulas~\cite{landau77,karnakov03} for the ground-state energy
of the hydrogen atom in a strong magnetic field. For both
asymptotic formulas a logarithmic-like behaviour $w_b/w_c\simeq
a\log_{10}(B/B_e)+b$, where $a$ and $b$ are positive constants,
has been determined.

The numerical estimate for the ratio of bound-state and
continuum-state decay rates has been performed. It has been found
that in contrast to the field-free case, where the bound-state
decay mode is suppressed by a factor of about
$4\times10^{-6}$~\cite{bahcall61,nemenov71,song87} as compared
with the usual (continuum-state) decay mode, in the presence of a
strong magnetic field $B\gtrsim B_e$ the ratio $w_b/w_c$ is of
the order $0.1\div0.4$. This remarkable finding can be important
for the physics of supernovae and neutron stars, where magnetic
fields with strength $B\gtrsim10^{13}$ G may exist. In
particular, the high value of the neutron bound-state decay rate
$w_b$ can influence the nucleon balance in protoneutron stars and
the hydrogen fraction in the atmosphere of neutron stars and
magnetars. The high value of $w_b$ can affect the neutrino
spectra from astrophysical objects with strong magnetic fields,
because in the neutron bound-state beta-decay process the energy
distribution of antineutrino is peaked about $\Delta-m_e$.
However, for estimating these possible astrophysical effects one
should consider a more involved problem, namely, the bound-state
beta-decay of a neutron moving in the presence of a strong
magnetic field and dense matter (for instance, such as a neutron
star). This implies modifications of the photon, electron, and
proton propagators in the Bethe-Salpeter
equation~(\ref{bethe-salpeter}) due to many-particle
effects~\cite{felipe91}. In addition, an electric field induced in
the rest frame of a neutron (due to its transverse motion in the
matter rest frame) should be accounted for, because for the
transverse neutron velocities $v_\perp\gg a_B|E^\parallel|$ the
induced electric fields are as strong as to pull the hydrogen
atom apart. Note, however, that in dense matter an electric field
can be strongly screened by the surrounding medium. The screening
also modifies the electron-proton interaction. In the context of
the present analysis the latter factor becomes appreciable if
$a_B\gtrsim l_s$, where $l_s$ is a screening length for the
medium. Since $l_s\sim n_c^{-1/3}$, where $n_c$ is a density of
charged particles (electrons or protons) in matter, we obtain the
following criterion $a_B\gtrsim n_c^{-1/3}$ (or $B\lesssim
n_c^{2/3}/e$) which indicates the situation where the role of
screening can not be neglected (such situation can be
encountered, for example, in the interiors of a protoneutron
star).

Let us remark that the high value of the bound-state decay rate
$w_b$ can be also important for cosmological applications. If the
controversial hypothesis of strong magnetic fields influencing
the usual beta-decay process in the early
Universe~\cite{grasso01} is realistic, then one has a right to
expect that the neutron beta-decay into a bound state of the
$(pe^-)$ system might have substantial consequences for big-bang
nucleosynthesis and the production of light elements in the early
Universe.

Note that in the terrestrial laboratory environment the strongest
magnetic field that can be produced is of the order of $10^7$~G
(see, for example,~\cite{crow98}) which is much lower than $B_e$.
However, the results of our present analysis can be used even in
the case of such fields if the neutron beta-decay takes place in
semiconducting or dielectric media, where a small effective mass
for the electron and a large dielectric constant reduce the
Coulomb force relative to the magnetic force~\cite{lai01}.

Finally, it is straightforward to generalize the results obtained
in this work to the bound-state beta-decay of a nucleus with
charge $Z-1$ in a strong magnetic field. This is realized by
replacing the nucleon mass defect $\Delta$ with the corresponding
value for the nucleus under consideration and taking into account
that the atomic energy and radius are given by
$\varepsilon=-Z^2e^4m_e$ and $a_0=(Ze^2m_e)^{-1}$, respectively.
After performing such a procedure, it can be deduced from
Eqs.~(\ref{decay_rate_free_2}),~(\ref{LL}) and/or~(\ref{KP}) that
$w_b/w_c\simeq Z[a\log_{10}(Z^{-2}B/B_e)+b]$, where the positive
constants $a$ and $b$ do not depend on $Z$. Thus, the dependence
of the ratio $w_b/w_c$ on the nuclear charge is different from
that in the absence of a magnetic field, in which case
$w_b/w_c\propto Z^3$.

\begin{acknowledgments}
We are very grateful to Yuri~V.~Popov for fruitful discussions on
the problem of neutron bound-state beta-decay. We are also
thankful to Vladimir~L.~Kauts for valuable comments.
\end{acknowledgments}
%


%

\eject

\begin{thebibliography}{99}
%
\bibitem{nemenov71}L.L.~Nemenov, Sov. J. Nucl. Phys. {\bf 15}, 582
(1972). 
%
\bibitem{bahcall61}J.N.~Bahcall, Phys. Rev. {\bf 124}, 495 (1961).
%
\bibitem{song87}X.~Song, J. Phys. G: Nucl. Phys. {\bf 13}, 1023
(1987).
%
\bibitem{fushiki92}I.~Fushiki, E.H.~Gudmundsson, C.J.~Pethick,
Astrophys. J. {\bf 342}, 958 (1989); T.A.~Mihara \textit{et al.},
Nature {\bf 346}, 250 (1990); G.~Chanmugam, Annu. Rev. Astron.
Astrophys. {\bf 30}, 143 (1992).
%
\bibitem{duncan92}R.C.~Duncan and C.~Thompson, Astrophys. J. {\bf
408}, L9 (1992); {\it ibid.} {\bf 473}, 322 (1996); V.V.~Usov,
Nature {\bf 357}, 472 (1992); M.~Baring and A.~Harding,
Astrophys. J. {\bf 304}, L55 (1998).
%
\bibitem{kulkarni98}S.R.~Kulkarni and C.~Thompson, Nature {\bf 393}, 215
(1998); C.~Kouveliotou {\it et al.}, {\it ibid.} {\bf 393}, 235
(1998).
%
\bibitem{grasso01}D.~Grasso and H.~Rubinstein, Phys. Rep. {\bf
348}, 163 (2001).
%
\bibitem{korovina64}L.~Korovina, Izvestia Vuzov. Fizika {\bf 6},
86 (1964); I.~Ternov, B.~Lysov, and L.~Korovina, Mosc. U. Phys.
Bull. {\bf 5}, 58 (1965); J.~Matese and R.~O'Connell, Phys. Rev.
{\bf 180}, 1289 (1969); L.~Fassio-Canuto, {\it ibid.} {\bf 187},
2141 (1969).
%
\bibitem{zakhartsev85}V.~Zakhartsev and Y.~Loskutov, Mosc. U.
Phys. Bull. {\bf 26}(2), 24 (1985); A.~Studenikin, Sov. J.
Astrophys. {\bf 28}(3), 639 (1988); A.~Studenikin, Sov. J. Nucl.
Phys. {\bf 49}, 1031 (1989).
%
\bibitem{bander92}M.~Bander and H.~Rubinstein, Phys. Lett. B {\bf
289}, 385 (1992).
%
\bibitem{shiff39}L.I.~Schiff and H.~Snyder, Phys. Rev. {\bf 55}, 59
(1939).
%
\bibitem{loudon59}R.~Loudon, Amer. J. Phys. {\bf 27}, 649 (1959).
%
\bibitem{eliott60}R.J.~Eliott and R.~Loudon, J. Phys. Chem. Sol.
{\bf 15}, 196 (1960).
%
\bibitem{hasegawa61}H.~Hasegawa and R.E.~Howard, J. Phys. Chem.
Sol. {\bf 21}, 179 (1961).
%
\bibitem{rau76}A.R.P.~Rau and L.~Spruch, Astrophys. J. {\bf 207},
671 (1976).
%
\bibitem{kashiev80}M.S.~Kaschiev, S.I.~Vinitsky, and F.R.~Vukajlovi\'{c},
Phys. Rev. A {\bf 22}, 557 (1980).
%
\bibitem{wang95}J.-H.~Wang and C.-S.~Hsue, Phys. Rev. A {\bf 52},
4508 (1995).
%
\bibitem{potekhin01}A.Y.~Potekhin and A.V.~Turbiner, Phys. Rev. A
{\bf 63}, 065402 (2001).
%
\bibitem{rutkowski03}A.~Rutkowski and A.~Poszwa, Phys. Rev. A {\bf 67},
013412 (2003).
%
%
\bibitem{karnakov03}B.M.~Karnakov and V.S.~Popov, JETP {\bf 97},
890 (2003).
%
%
\bibitem{lindgren79}K.A.U.~Lindgren and J.T.~Virtamo, J. Phys. B:
At. Mol. Phys. {\bf 12}, 3465 (1979).
%
\bibitem{goldman91}S.P.~Goldman and Z.~Chen , Phys. Rev. Lett. {\bf 67},
1403 (1991).
%
\bibitem{chen93}Z.~Chen and S.P.~Goldman, Phys. Rev. A {\bf 48},
1107 (1993).
%
\bibitem{doman80}B.G.S.~Doman, J. Phys. B:
At. Mol. Phys. {\bf 13}, 3335 (1980).
%
\bibitem{herold81}H.~Herold, H.~Ruder, and G.~Wunner,
J. Phys. B: At. Mol. Phys. {\bf 14}, 751 (1980).
%
\bibitem{potekhin94_98}A.Y.~Potekhin, J. Phys. B: At. Mol. Opt.
Phys. {\bf 27}, 1073 (1994); {\bf 31}, 49 (1998).
%
\bibitem{karplus52}E.E.~Salpeter, Phys. Rev. {\bf 87}, 328 (1952);
R.~Karplus and A.~Klein, {\it ibid.} {\bf 87}, 848 (1952).
%
\bibitem{shabad86}A.E.~Shabad and V.V.~Usov, Astrophys. Space
Science {\bf 128}, 377 (1986).
%
\bibitem{sokolov68}A.A.~Sokolov and I.M.~Ternov, {\it Synchrotron
Radiation} (Pergamon Press, New York, 1968).
%
\bibitem{abramowitz72}M. Abramowitz and I.A. Stegun (Eds.),
{\it Handbook of Mathematical Functions, 9th printing} (Dover, New
York, 1972).
%
\bibitem{lai01}D. Lai, Rev. Mod. Phys. {\bf 73}, 629 (2001).
%
\bibitem{TerBagDorJETP69}I.~Ternov, V.~Bagrov, V.~Bordovitsyn,
and O.~Dorofeev, JETP {\bf 28} 1206 (1969).
%
\bibitem{Schw88}J.~Schwinger, {\it Particles, sources and fields, vol.3} (Addison-Wesly, Redwood City, CA, 1989).
%
\bibitem{leinson85}L.B.~Leinson and V.N.~Oraevsky, Sov. J. Nucl.
Phys. {\bf 42}, 254 (1985).
%
\bibitem{landau77}L.D.~Landau and E.M.~Lifshitz, {\it Quantum Mechanics, 3rd
edition} (Pergamon, Oxford, 1977).
%
\bibitem{felipe91}R.~Gonzalez Felipe, A.~Perez Martinez, and A.E.~Shabad, Phys. Rev. A {\bf 43}, 5575 (1991).
%
\bibitem{crow98}J.E.~Crow, J.R.~Sabin, and N.S.~Sullivan, in {\it Atoms and
Molecules in Strong Magnetic Fields}, edited by P. Schmelcher and
W. Schweizer (Plenum, New York, 1998), p. 77.
%
\end{thebibliography}
\end{document}